\documentclass[fleqn,12pt,twoside]{article}
\usepackage{espcrc1}

\usepackage{graphicx}
\usepackage[figuresright]{rotating}

\def\be{\begin{eqnarray}}
\def\ee{\end{eqnarray}}
\def\ben{\begin{enumerate}}\def\bitem{\begin{itemize}}
\def\een{\end{enumerate}}\def\eitem{\end{itemize}}
\def\bi{\bibitem}

\def\Kp{$K^+$}\def\Km{$K^-$}

\def\pr{Phys. Rev.}
\def\pl{Phys. Lett.}

\def\roughly#1{\mathrel{\raise.3ex\hbox{$#1$\kern-.75em%
\lower1ex\hbox{$\sim$}}}}

\def\MeV{{\mbox{MeV}}}

\newcommand{\AmS}{{\protect\the\textfont2
  A\kern-.1667em\lower.5ex\hbox{M}\kern-.125emS}}

\title{Broad Band Equilibration of Strangeness}

\author{Gerald. E. Brown\address[sunysb]{Department of Physics and Astronomy, 
        State University of New York\\ 
        Stony Brook, NY 11794-3800, USA},
        Mannque Rho\address{Service de Physique Th\'eorique, CE Saclay, 91191 
        {\it Gif-sur-Yvette}, France} 
        and
        Chaejun Song\addressmark[sunysb]\thanks{GEB and CS were supported by 
the US Department of Energy under Grant No. DE-FG02-88 ER40388.}
        }
\begin{document}

\maketitle

\begin{abstract}
We develop the ``broad band equilibration" scenario for kaon productions
at GSI energies with in-medium effects. 
\end{abstract}

\section{Introduction}

As shown in some talks~\cite{stat} of this conference, 
statistical descriptions work very
successfully for  
multihadron final states in nucleon-nucleon, nucleon-nucleus, and
nucleus-nucleus collisions. In this approach the chemical equilibrium 
particle ratios are functions of temperature $T$ and baryon  
chemical potential $\mu_{B}$ only\footnote{In the canonical description
for heavy ion collision we need a system size parameter but the
parameters are related to the number of participants. For example,
$V\sim 1.9\pi A_{p}$~\cite{cor98}. And in the grand canonical description
we need the chemical potentials of conserved charges.}.

\begin{figure}
\centerline{\includegraphics[scale=0.3]{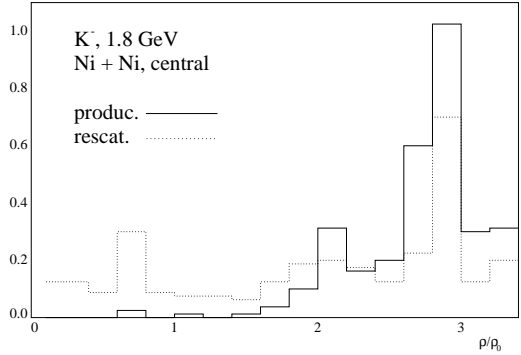}}
\caption{\small
Bratkovskaya and Cassing's calculations of \Km production
(private communication). The solid line and dashed line give the 
density of the origin and that of the last interaction, respectively.}
\label{bc}
\end{figure}

Recently GSI measured Au-Au and Ni-Ni collisions at 0.8 $\sim$ 2.0 GeV.
Some measured particle multiplicity ratios are in Table \ref{table:1} below.
\begin{table}[htb]
\caption{\small Experimental results for different particle ratio in 
central Ni $+$ Ni collisions.
}
\label{table:1}
\newcommand{\m}{\hphantom{$-$}}
\newcommand{\cc}[1]{\multicolumn{1}{c}{#1}}
\renewcommand{\tabcolsep}{2pc} 
\renewcommand{\arraystretch}{1.2} 
\begin{tabular}{lllll}
\hline
$\pi^+/p$ & $K^+/\pi^+$ & $\pi^-/\pi^+$ & $d/p$ & $K^+/K^-$ \\
\hline
\m0.17  & \m0.0084 & \m1.05 & \m0.28 & \m32 \\
\hline
\end{tabular}\\[2pt]
All the data except for $K^+/K^-$
comes from \cite{cor98} at $E=1.8$ A GeV. $K^+/K^-$ comes
from \cite{menzel} at $E=1.93$ A GeV.
\end{table}
Cleymans, Oeschler and Redlich~\cite{cor98} fit them well by
a common chemical freezeout at $T=70$ MeV and $\mu_B=750$ MeV
in terms of a canonical thermal model. This is another success
of statistical description. However, it is hard to understand
that the chemical freezeout density from these values of $T$
and $\mu_B$ is just $\rho_0/4$ where the \Km mean free path is
$\sim$ 6 fm. 
Bratkovskaya and Cassing's transport calculations 
in Fig.~\ref{bc} show that \Km production comes from all
densities.

We can understand simply why they are compelled to obtain such a
low density. By strangeness conservation the number of \Kp is 
the sum of the number of \Km and that of $\Lambda$\footnote{Of course
we need to include $\Sigma$ and $\Xi$ hyperon. Here they are
neglected to simplify the discussion.} We measured \Kp/\Km $\sim$ 32 in 
Table~\ref{table:1} so $\Lambda$/\Km $\sim$ 31. Thus if we get 
the right $\Lambda$/\Km ratio, \Kp/\Km follows. The ratio is
\be
R=\frac{\Lambda}{K^-}
=\left(\frac{\bar{p}_\Lambda}{\bar{p}_{\bar{K}}}\right)^{\frac32}
\frac{e^{-(E_\Lambda -\mu_B)/T}}{e^{-E_{\bar{K}}/T}}
\approx \left( \frac{M_\Lambda}{M_{\bar{K}}} \right)^{\frac32}
\frac{e^{-(M_\Lambda -\mu_B)/T}}{e^{-M_{\bar{K}}/T}}.\label{ratio}
\ee
If $\mu_B=M_N$, $M_\Lambda =M_N+ 175$ MeV gives $R\sim $ 280. It's too
large so they need $\mu_B=750$ MeV, 190 MeV lower than $M_N$,
in order to get the right value of $R$. That's why they obtained the low
density for equilibration with free-space mass hadrons.

\section{In-medium kaons}
\begin{figure}
\centerline{\includegraphics[scale=0.8]{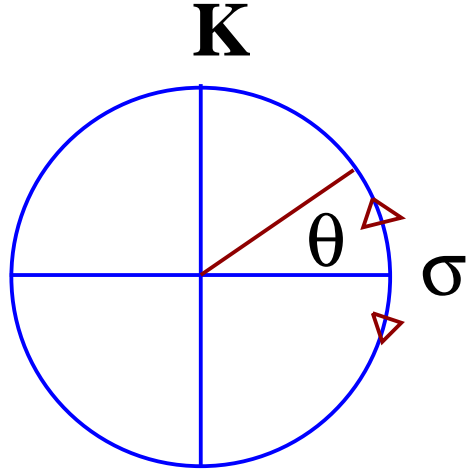}}
\caption{\small V-formalism}
\label{vform}
\end{figure}

A simple way to introduce an in-medium \Km mass is
the V-spin formalism in \cite{BKR87}. In the formalism we can 
obtain the kaon effective mass by considering the kaon 
fields as small fluctuations 
$\theta$ around the $\sigma$ axis as shown in Fig.\ \ref{vform}. 
The Hamiltonian for explicit
chiral symmetry breaking is 
\be
H_{X\chi SB}=\Sigma_{KN}<\bar{N}N>\cos\theta
+\frac12 m_K^2f^{\star 2}\sin^2\theta.
\ee
Then the effective mass of kaons (small fluctuation $\theta$) comes to 
drop from movement towards restoration of explicitly broken chiral
symmetry;
\be
\frac{m^{\star 2}_K}{m^2_K}=1-\frac{\Sigma_{KN}\rho}{f^{\star 2} m_K^2}
\ee
with an approximation $<\bar{N}N>\approx\rho$. Taking $m_s\sim 150$
MeV, $m_u=m_d=6$ MeV, $\Sigma_{\pi N}=46$ MeV and $<N|\bar{s}s|N>
\sim\frac13<N|\bar{d}d|N>$~\cite{liu}, we obtain
\be
\Sigma_{KN}=\frac{(m_u+m_s)<N|\bar{u}u+\bar{s}s|N>}{(m_u+m_d)
<N|\bar{u}u+\bar{d}d|N>}\Sigma_{\pi N}\sim 400\ \MeV .
\ee
The effective value of $\Sigma_{KN}$ is somewhat smaller by range 
corrections~\cite{chl};
\be
(\Sigma_{KN})_{eff}=(1-0.37\omega_K^{\star 2}/m_K^2 )\Sigma_{KN}
\ee
where $\omega^\star_{K^-}=m_{K^-}-\frac{\omega^\star_{K^-}}{m_K}V_K$
is the self-energy of the antikaon at rest with vector potential $V_K$. 

\section{Broad equilibration and kaon condensation}

Considering kaon mass shift in medium, we can rewrite $R$ in Eq.~(\ref{ratio}) 
as
\be
R=\left( \frac{M_\Lambda}{m^\star_{K^-}}\right)^{3/2}
e^{(\mu_B+\omega^\star_{K^-})/T}e^{-M_\Lambda /T}.
\ee
We assume that kaon vector potential $V_K$ is proportional 
to the density below $\rho_0$ and constant above $\rho_0$, because 
the vector decoupling is expected at high density~\cite{vec}. 
As $\mu_B$ goes from 860 MeV to 905 MeV, $\rho$ goes from 1.2$\rho_0$
to 2.1$\rho_0$ and $\mu_B+\omega^\star_{K^-}$ goes from 1230
to 1227 MeV with fixed $T=$ 70 MeV. Even at $\rho_0/4$
$\mu_B+\omega^\star_{K^-}$ is 1245 MeV.
Disregarding the dependence on $m^\star_{K^-}$ in the denominator
and on T\footnote{The increase of $R$ by $m^\star_{K^-}$ in the denominator
can be compensated for by increasing $T$.} for simplicity,
we found the ratio $R$ is independent of density.
This results can be checked by the recent experimental results. 
The ratio \Kp/\Km ,which is directly related to $\Lambda$/\Km ,
is almost independent of centrality in GSI~\cite{menzel} and AGS~\cite{dunlop}
experiments.
In other words, the ratio of hyperon to \Km is roughly the
same at all densities. We call it ``broad-band equilibration". 

So the negatively charged strangeness sector looks
well and truly equilibrated within itself.
Positively charged strangeness hardly changes as the system expands
below 2$\rho_0$~\cite{BKWX}. Since the number of positively charged
strangeness is the same as that of the negatively charged strangensess,
most of the negatively charged strangeness particles will be produced
early.

The strong attraction that the \Km experiences in dense matter can make
the kaon energy come down to the electron chemical potential and 
kaons be able to replace electrons. If the vector decouples in the changeover
from nucleons to constituent quarks as variables as we expect,
the kaon will condensate 
before chiral restoration.
The kaon condensation sets the maximum 
neutron star mass 1.5 $M_\odot$~\cite{llb}.
This is consistent with the observation that all the neutron
star masses are below 1.5 $M_\odot$, in systems with degenerate
companions, where the companion structure should not influence
the measurement.

\end{document}